\documentclass[twoside,a4paper,11pt]{sca}
\def\aj{AJ}%
\def\araa{ARA\&A}%
\def\apj{ApJ}%
\def\apjl{ApJ}%
\def\apjs{ApJS}%
\def\apss{Ap\&SS}%
\def\aap{A\&A}%
\def\mnras{MNRAS}%

\def\msun{\hbox{M$_\odot$}}

\def\t4{\hbox{t$_{\rm 4}$}}
\def\fmid{\hbox{F$_{\rm MID}$}}

\usepackage{graphicx}
\usepackage{hyperref}
\usepackage{movie15}
\usepackage{natbib}  
\begin{document}
\pagenumbering{arabic}
\pagestyle{myheadings}
\thispagestyle{empty}
{\flushright\includegraphics[width=\textwidth,bb=90 650 520 700]{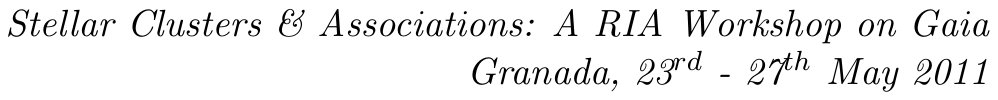}}
\vspace*{0.2cm}
\begin{flushleft}
{\bf {\LARGE
%
Cluster Disruption: From infant mortality to long term survival%
}\\
\vspace*{1cm}
%
Nate Bastian$^{1}$
%
}\\
\vspace*{0.5cm}
%
$^{1}$
Excellence Cluster Universe, Boltzmannstr. 2, 85748 Garching, Germany\\
%
\end{flushleft}
%
\markboth{
Cluster Disruption
}{ 
%
N. Bastian
%
}
\thispagestyle{empty}
\vspace*{0.4cm}
\begin{minipage}[l]{0.09\textwidth}
\ 
\end{minipage}
\begin{minipage}[r]{0.9\textwidth}
\vspace{1cm}
\section*{Abstract}{\small
%
How stellar clusters disrupt, and over what timescales, is intimately linked with how they form.  Here, we review the theory and observations of cluster disruption, both the suggested initial rapid dissolution phase (infant mortality) and the longer timescale processes that affect clusters after they emerge from their progenitor GMCs.  Over the past decade, the standard paradigm that has developed is that all/most stars are formed in clusters and that the vast majority of these groups are disrupted over short timescales ($<10$ Myr).  This is thought to be due to the removal of the left over gas from the star-formation process, known as infant mortality.  However, recent results have suggested that the fraction of stars that form in clusters has been overestimated, with the majority being formed in unbound groups (i.e. associations) which expand and disrupt without the need of invoking gas removal.  Dynamical measurements of young massive clusters in the Galaxy suggest that clusters reach a stable equilibrium at very young ($<3$~Myr) ages, suggesting that gas expulsion has little effect on the cluster.  After the early dynamical phase, clusters appear to be long lived and stable objects. We use the recent WFC3 image of the cluster population in M83 to test empirical disruption laws and find that the lifetime of clusters strongly depends on their ambient environment.  While the role of cluster mass is less well constrained (due to the added parameter of the form of the cluster mass function), we find evidence suggesting that higher mass clusters survive longer, and that the cluster mass function (at least in M83, outside the nuclear region) is truncated above $\sim10^5$\msun.

%
\normalsize}
\end{minipage}
%
%
%

\section{The Standard Paradigm of Early Cluster Evolution and Disruption}

Star clusters form in the dense cores of giant molecular clouds.  Assuming a non-100\% star-formation efficiency (SFE; i.e., that only some fraction of the gas will turn into stars), at some point a natal cluster will be made of both gas and stars, and both will contribute to the total gravitational potential.  Due to stellar feedback, thought to come mainly from the winds and photoionisation of massive stars (as clusters appear to be largely devoid of gas before the first supernovae occurs), the gas may be rapidly removed from the young cluster.  If the system was in virial equilibrium  the removal of the gas will leave the stellar part of the system out of equilibrium, in a super-virial state, and the stars will expand in order to find a new equilibrium (c.f., Tutukov 1978; Hills~1980).  This will result in the loss of some stars from the system, and if the SFE is less than 33\% the entire cluster may become unbound (assuming instantaneous gas removal), whereas above this critical threshold, some bound core may survive (e.g., Bastian \& Goodwin~2006).

This is a conceptually simple theory, with well defined initial conditions that can be easily parameterised.  This allows for large systematic theoretical and numerical studies to be undertaken in order to find out how each parameter affects the final state of the remnant cluster (if such a remnant exists).  For the 'classical' models (i.e. those that assume the system is in equilibrium at the beginning of the simulation - see Lada et al.~1984), the main parameters are: 1) the SFE, i.e., the fraction of gas turned into stars (e.g., Hills~1980); 2) the timescale over which the gas is removed (e.g., Goodwin~1997; Baumgardt \& Kroupa~2007) and 3) how the gas and stars are distributed within the system, i.e. if they have the same profile shape/size (e.g., Chen \& Ko~2009).    Generally, the system is modelled by adopting an analytic background potential for the gas and using N-body techniques for the stars.

Additionally, if full cluster populations are modelled, then one can also make assumptions on how each of the three above parameters vary with other cluster properties, such as mass (e.g, Kroupa \& Boily~2002) or on  the initial conditions of the population (e.g., any primordial mass-size relations; Parmentier \& Kroupa 2011).  The response of the stellar system to gas removal is largely independent of the number of stars in the system (i.e. stellar mass) as it is driven by violent relaxation (e.g., Goodwin \& Bastian~2006).

A new flavour of models has recently been explored which relaxes the assumption of virialized starting conditions.  In these models, like in the 'classical' case, the gas is modelled as a background potential, however the stars are either in a sub- or super-virial state with clumpy initial conditions (e.g., McMillan et al.~2007, Allison et al. 2009).  For the sub-virial case, i.e. the stars are moving slower than in the equilibrium case, the stars fall toward the center of the system, while the gas remains in the initial configuration.  Such sub-virial motions have been found in turbulent hydrodynamical simulations (Offner et al.~2009) and many young star forming regions appear to be highly sub-structured (see \S~\ref{sec:substructure}), hence the initial conditions of young clusters may well be sub-virial and clumpy.  Sub-virial motions result in an 'effective' SFE (eSFE; c.f., Goodwin~2009) for the central regions that is significantly higher than the mean SFE (note that this is similar to assuming different profile shapes/sizes for the star and gas distributions).    This contraction of the stellar system can largely offset the effects of gas expulsion (e.g., Smith et al.~2011).  For the super-virial case, the effect of gas expulsion is amplified.

The end result of the simulations presented to date is that a bound cluster is more likely to form if: 1) the (e)SFE is high; 2) the timescale of gas removal is long compared to a crossing time of the system; or 3) the system begins in a sub-virial state.  In recent SPH simulations of star cluster formation, the sub-clumps that merge in order to form a cluster appear to have extremely high eSFE (Moeckel \& Clarke~2011; Kruijssen et al.~2011b), which limits (or even negates) the effect of gas expulsion on the system.

In the next section we discuss the observations that the infant mortality paradigm was designed to explain, and review ongoing studies to see if such a scenario is still necessary.  

\section{Initial Conditions}

\subsection{Do all stars form in clusters?}
\label{sec:substructure}

Star-formation appears to be hierarchical within galaxies, meaning that there does not seem to be a preferred scale in the star-formation process, from $\sim0.1$ pc to hundred parsec scales (e.g. Elmegreen et al.~2006; Bastian et al.~2007; 2009; S{\'a}nchez et al.~2010).  This is likely simply due to the fact that the ISM appears hierarchical (e.g. Elmegreen \& Falgarone~1996) and that star-formation passively traces the gas.  Because of this initial spatial distribution we would not expect to see all stars forming in clusters, but instead would expect a continuous distribution of surface densities ranging from the dense inner cores of massive clusters to extremely isolated stars.  Pre-{\it Spitzer} studies that attempted to place constraints on the spatial distribution of young stellar objects (YSOs) used K-band excess techniques (i.e. looking for overdensities in K-band images) which were excellent in picking up the dense inner structures of clusters, but largely missed the lower density regions (e.g., Carpenter~2000).  This observational bias, led to the notion that all stars form in clusters.

\begin{figure}
\center
\includegraphics[width=9.7cm,angle=0,clip=true]{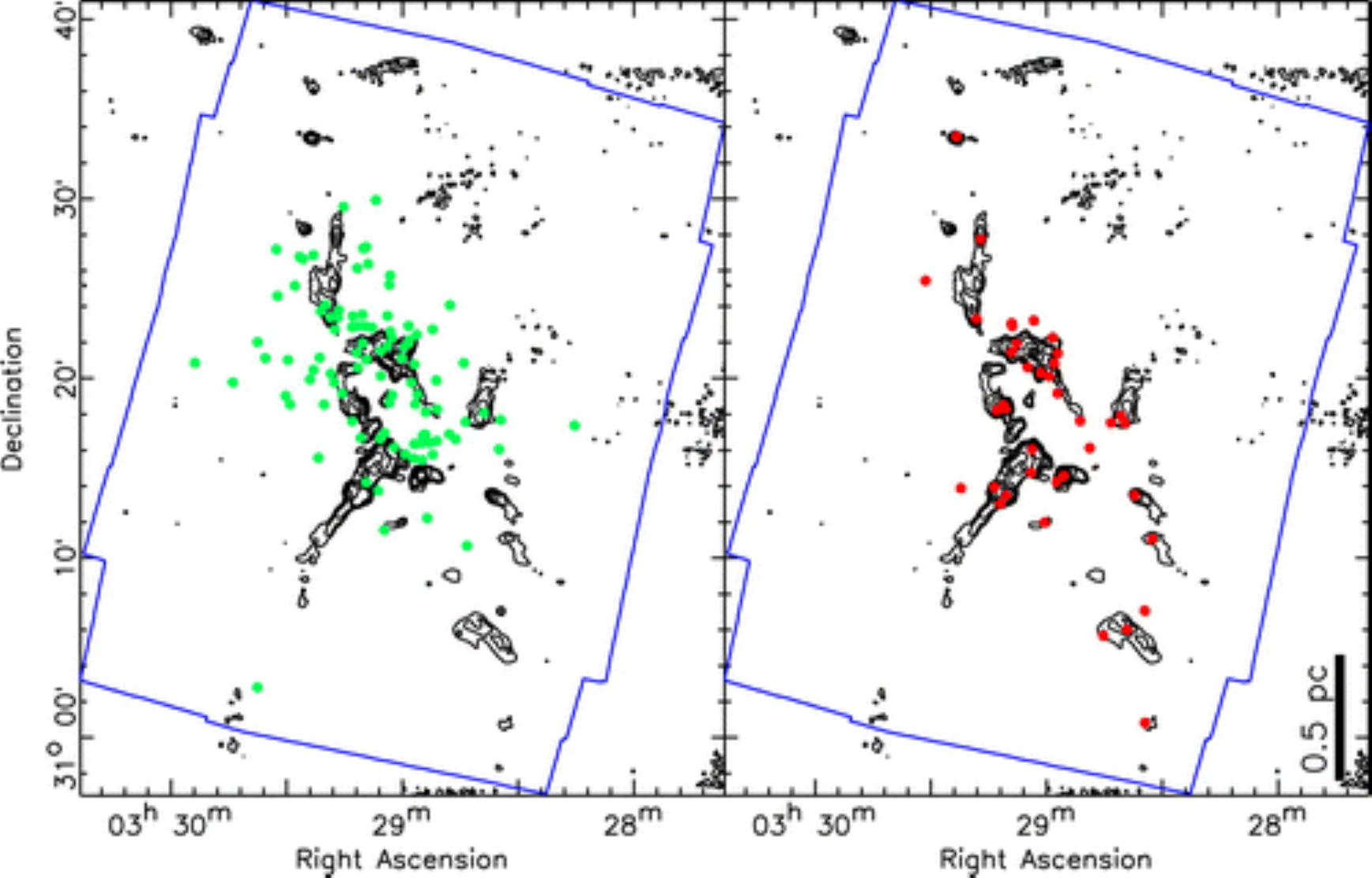} 
\caption{\label{fig:gutermuth} (from Gutermuth et al.~2008) The spatial distribution of YSOs in the young embedded star-forming region NGC~1333.  The contours show dense gas and the points represent Class Is (red, right panel) and Class IIs (green, left panel).  Class Is are thought to be younger, on average, and their spatial distribution matches the gas distribution quite well, showing a very filamentary structure.  The Class IIs also generally follow the gas, however, due to their older age they have dynamically evolved to some degree.  This shows that stellar clusters do not form as a centrally concentrated system with the gas/stars in equilibrium, but rather as a filamentary/hierarchical structure where individual components may (or may not) merge to form the final stellar system.
}
\end{figure}

However, with the advent of the {\it Spitzer Space Telescope}, YSOs could be identified based on the mid-IR colours, i.e. independent of their surface density.  In Fig.~\ref{fig:gutermuth} we show the spatial distribution of YSOs in the star-forming region, NGC~1333 (Gutermuth et al.~2008).  Note their highly filamentary structure and the fact that they trace the gas distribution quite closely.  This is a fairly typical example of a star-forming region, showing multiple sub-clumps which may or may not merge as the cluster evolves dynamically (e.g., Gutermuth et al.~2009; Schmeja et al.~2009).  The hierarchical nature of star-forming regions becomes more clear when one zooms out an order of magnitude in spatial scale and looks at the structure of full clouds, with nested clumps from sub-parsec to tens of parsec scales (c.f.~Elmegreen et al.~2000; Allen et al.~2007).

In Fig.~\ref{fig:bressert} we show the surface density distribution of YSOs for a nearly complete sample of all star-forming regions within 500~pc as determined through {\it Spitzer} imaging (Bressert et al.~2010).  Note that there does not appear to be any preferred scale, i.e. no way to distinguish between clusters, associations, or distributed star-formation.   The vertical lines in the bottom panel refer to previous definitions that had been applied in the literature, resulting in a large range of estimates for the fraction of star-formation in clusters, from 90\% to $<25\%$.

A similar conclusion was also reached by Gieles \& Portegies Zwart~(2011; also see the contribution of M. Gieles in these proceedings) who found a continuous distribution, from loose associations to dense clusters, in the structures of star-forming to evolved stellar groups.

\subsection{Implications for infant mortality}
\label{sec:im}

\begin{figure}
\center
\includegraphics[width=8.3cm,angle=0,clip=true]{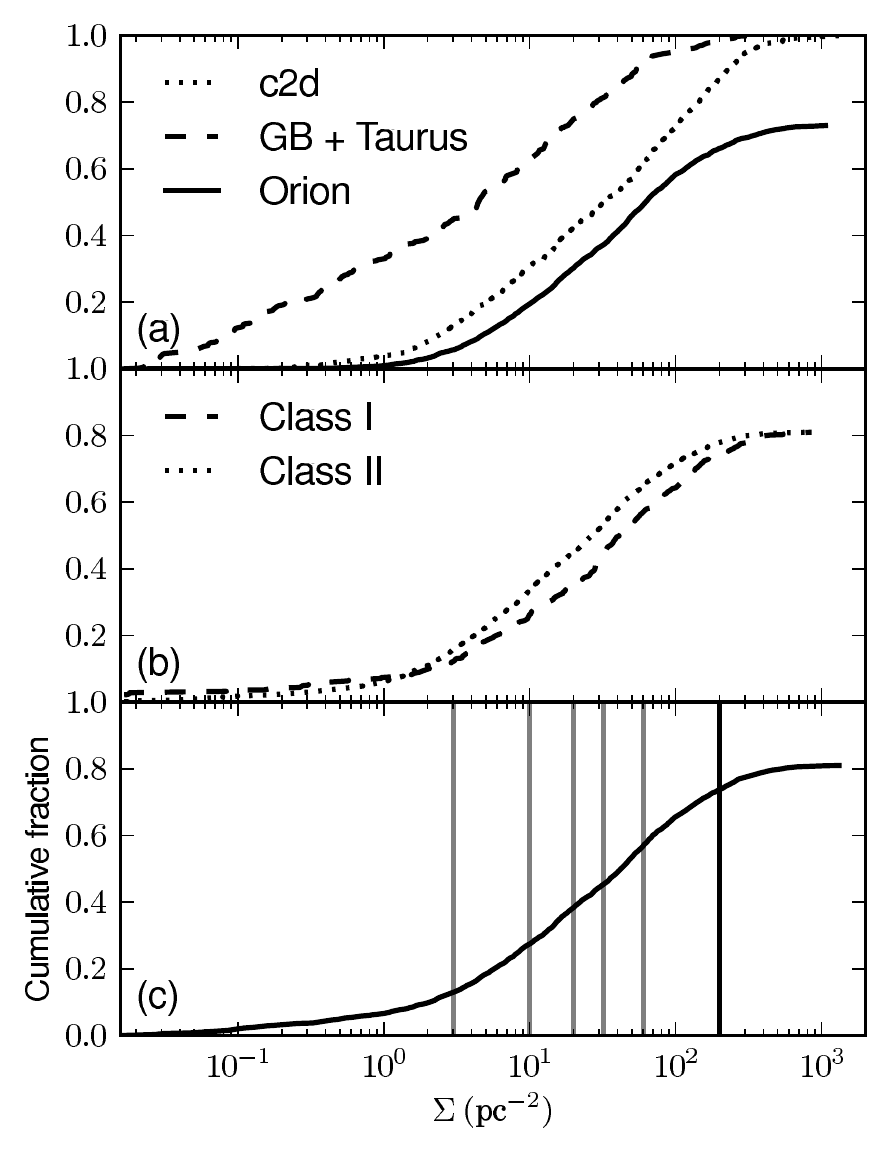} 
\caption{\label{fig:bressert} (from Bressert et al.~2010)(a) The cumulative fraction of surface densities for the Gould Belt and Taurus (GB+Taurus), Cores to Disks (c2d), and Orion surveys. Each SF region included in the distributions has N(YSOs) $\geq 10$ and a sufficient field-of-view to properly calculate stellar surface densities. The Orion survey stops at 73\% for the cumulative fraction since the ONC is excluded. A  65\% disk fraction for all of the SF regions is adopted and each curve is normalised by the number of YSOs in each survey. (b) With the GB+Taurus, c2d, and Orion surveys combined we see Class I \& II distributions having similar profiles with a small off- set in density, showing that we are likely seeing the primordial distribution of the YSOs. (c) With all of the Spitzer surveys combined we compare several cluster definitions. The vertical grey lines from left to right are Lada \& Lada (2003), Megeath et al. (in prep.), J{\o}rgensen et al. (2008), Carpenter (2000), and Gutermuth et al. (2009) stellar density requirements for clusters. These values correspond to 3, 10, 20, 32, and 60 YSOs pc$^2$ and intersect the corrected cumulative distribution profile, implying that 87\%, 73\%, 62\%, 55\%, and 43\% of stars form in clusters, respectively. The percentages correlate to what fraction of stars form in ÔclustersÕ based on the various definitions. The black vertical line is for a dense environment where $\Sigma \geq 200$ YSOs/pc$^2$. The fraction of YSOs in a dense environment is $< 26$\%.
}
\end{figure}

Lada \& Lada~(2003) compared the number of observed open clusters with that expected given their embedded cluster sample (see Fig.~\ref{fig:bressert} for the definition that they applied to define an embedded cluster).  They found that they would expect $\sim10$ times more open clusters than observed if all embedded clusters evolve into open clusters.  Due to this large difference, they postulated that $\sim90$\% of clusters are disrupted as they pass from an embedded to an exposed state.  The authors named this mass culling, ``infant mortality".  The suspected cause of this disruption was the removal of the left-over gas from the star-formation process, i.e. rapid gas removal (e.g. Geyer \& Burkert 2001; Kroupa \& Boily~2002; Bastian \& Goodwin~2006).  

However, how much infant mortality is required depends strongly on how one defines an embedded cluster.  If more conservative definitions are used (see Fig.~\ref{fig:bressert}) instead of 90\% of all clusters being destroyed during this phase (or equivalently, due to the cluster mass function, each cluster losing 90\% of its mass), less than 50\% of clusters need to be destroyed in order to explain the observed open cluster numbers.  Since it is difficult or impossible to define which stars are part of an embedded cluster, hence the embedded cluster's mass, other techniques are required to test the infant mortality scenario.

How important is gas expulsion?  This is difficult to address using populations, due to the uncertainties outlined above.  However, one can address this through detailed dynamical studies of individual clusters.  Goodwin \& Bastian~(2006) showed that the observed dynamical mass (obtained through measuring the velocity dispersion and size of a cluster) of a cluster should be larger than the actual mass (obtained through comparing the observed light with M/L ratios from SSP models) if the cluster is undergoing expansion due to gas expulsion.  The authors also presented a series of observations from the literature that showed that young clusters had systematically larger dynamical mass estimates, consistent with the idea of gas expulsion.  However, Gieles et al.~(2010) showed that this effect could be mimicked by the broadening of the velocity dispersion due to binary stars.  Since O and B stars are more likely to be in tight binary systems than their lower mass counterparts, and that these massive stars dominate the light of young massive clusters, this broadening of the velocity dispersion will affect young clusters more than older clusters.  This can also explain the observations without invoking infant mortality.   Hence, studies of unresolved extragalactic clusters are not able to address the role of gas expulsion.

In order to make progress, multi-epoch high-resolution spectroscopy of individual cluster member stars is required, which is only possible in the Galaxy and SMC/LMC.  The multi-epoch nature allows binaries to be detected and removed from the sample, resulting in a clean velocity dispersion estimate.  Alternatively, proper motion studies of massive clusters are also not affected by binaries.  While this is difficult and time consuming work, a handful of studies have been carried out to date.  Studies of NGC~3603 (Rochau et al.~2010), Westerlund 1 (Mengel \& Tacconi-Garman~2007, Cottaar et al. in prep, Cottaar these proceedings) the Arches (Clarkson et al.~2011; Clarkson et al. in prep) and R136 in the LMC (H{\'e}nault-Brunet et al. in prep) have all found low velocity dispersions for these young clusters.  This is particularly surprising for NGC~3603 and the Arches, given their very young ages ($\leq 2$~Myr; see also the contribution of M. Gieles in these proceedings).  

{\em These results suggest that gas expulsion has not significantly affected these clusters and that once a cluster forms it is stable from a very young age.  Hence, infant mortality may not be an important process in the early stages of cluster evolution}.  Whether the cluster stability comes from a high star-formation efficiency (e.g. Goodwin \& Bastian~2006), the merging of sub-virial sub-clumps (e.g. Allison et al.~2009) or through a de-coupling of the gas and stars in a highly dynamical process (e.g. Moeckel \& Clarke~2011; Kruijssen et al.~2011b; see also Kruijssen in these proceedings) is uncertain at present.

\subsection{Why it matters}

In addition to the question of whether infant mortality and gas expulsion are important phases in young cluster evolution, whether or not all stars form in clusters is important for a number of other studies.  One issue that {\it GAIA} will make an important contribution to, is the question of whether associations are merely expanded clusters (due to gas removal), or if associations are merely part of the continuous hierarchical distribution of star-forming structures.  {\it GAIA} will measure the proper motions of thousands of stars in young clusters and associations in the Galaxy.  If all stars form in clusters, then the stars in associations should be traced back to a handful of specific points (i.e., the location of their parent clusters at the moment of gas expulsion).  On the other hand, if associations are simply a level in a continuous hierarchy, the stars should have (largely) random motions.

If all stars form in clusters, then clusters can be viewed as a fundament unit in the star forming process.  In this view, star formation can be broken up into discrete, or quantised events.  Since the difference in age between two stellar groups (i.e. two sub-clumps within a young cluster) is related to their separation distance, with spatially close groups having similar ages (Efremov \& Elmegreen~1998), young clusters (with radii less than a few parsecs) are single age populations to a good approximation.  Alternatively, if many stars are born in associations, which span 10s or 100s of parsecs, then the individual sub-groups may have significantly different ages, hence star formation would not be made up of discrete events.

Whether or not star formation happens in a continuous/hierarchical way or is made up of discreet events has important implications for the integrated properties of stellar populations (see the contributions by E. Bressert and C. Weidner in these proceedings).  For example, Weidner \& Kroupa~(2006) have suggested that the mass of the most massive star and the mass of the host cluster have a causal relation, meaning that a cluster {\it must} have a minimum mass before it forms massive stars.  However, this only works if clusters are a basic unit in the star-formation process, as this relation can not hold for the individual sub-clumps that make a cluster {\it and} the final cluster at the same time.  Nor could the relation hold for larger associations/complexes of clusters.  Hence there would need to be a preferred scale in the process, i.e., a star must know about the final mass of the cluster that it will end up in.  This type of behaviour has been seen in SPH simulations of cluster formation based on competitive accretion (Maschberger et al.~2010), where the masses of stars are determined in large part by the environment from which they form.

Additionally, the spatial distribution of star-formation can provide constraints on theories of star (and cluster) formation (e.g., Bressert et al.~2011).  Cluster disruption, specifically the infant mortality phase, is the largest uncertainty is addressing the role of environment in the cluster formation process, i.e., if the fraction of stars formed in clusters is related to the star-formation rate (or SFR surface density) of the host galaxies (e.g., Goddard et al.~2010; Silva-Villa \& Larsen~2011; Adamo et al.~2011).




\section{Further Evolution}

After a cluster transitions from an embedded to an exposed state, the surviving clusters are not expected to live indefinitely, but rather to disrupt due to internal (e.g., two-body relaxation, stellar evolution) and external (e.g., tidal fields and interactions with GMCs) processes.  Fall, Chandar \& Whitmore~(2009) have suggested that these processes happen over different timescales, and hence can be treated independently, in addition to being largely independent of cluster mass.  Alternatively, through an analytic approach, Gieles et al.~(2011) find that all processes are acting concurrently and that the lifetime of a cluster depends on the ambient environment and its mass, in agreement with other theoretical investigations (Spitzer 1958, H{\'e}non 1961, etc).

Observationally, the situation is less clear, and the community has yet to reach a consensus regarding the amount of disruption observed in populations as well as the role of cluster mass and environment in the process.  There have been two main empirical disruption laws put forward in the literature, based on different cluster samples in a variety of galaxies (explained in detail below).  However, in only one case, the SMC, have the advocates for the different scenarios used the same dataset (Chandar, Fall \& Whitmore~2006; Gieles, Portegies Zwart \& Lamers~2007), and even here the authors come to different conclusions.  

Below we summarise the two main empirical disruption laws and use the new WFC3 imaging of the cluster population of the spiral galaxy M83 to test the scenarios.

\subsection{Empirical disruption laws}

\subsubsection{Mass dependent disruption (MDD)}

The first empirical disruption law considered here is {\it Mass Dependent Disruption} (MDD).  This was first presented, in the form used here, in Boutloukos \& Lamers~(2003), in order to explain the observed cluster population properties in the Galaxy, SMC, M33 and M51 (however, see Larsen~(2008) for a thorough review of empirical mass dependent disruption laws).  In this scenario, the lifetime of a cluster is dependent on the initial mass of the cluster as $M^{\gamma}$ with $\gamma \sim 0.62$, and on the ambient environment with clusters surviving longer in galaxies with weak tidal forces and low numbers of GMCs.  The empirical model was updated in Lamers et al.~(2005), applied to a distance limited sample of open clusters in the Milky Way in Lamers \& Gieles~(2006), and was shown to agree with predictions from numerical $N$-body experiments in Gieles et al.~(2004).

\subsubsection{Mass independent disruption (MID)}

The second empirical disruption law considered here is {\it Mass Independent Disruption} (MID).  In this scenario, cluster disruption is independent of the cluster mass and the ambient environment.  While the classic infant mortality falls in this category and is thought to last for $\leq10$~Myr (e.g. Lada \& Lada~2003), the concept has been expanded up to ages of $\sim1$~Gyr and has been invoked to explain cluster populations in the Antennae (Fall et al.~2005), the LMC (Chandar et al. 2010a), and M83 (Chandar et al. 2010b).  The model is described in detail in Whitmore et al.~(2007). The MID scenario suggests that 90\% of clusters are disrupted every decade in age, independent of their mass or the environment that the cluster forms/evolves in.  Hence, if 1000 clusters are formed at a given time, 100, 10 and 1 will be left after 10~Myr, 100~Myr and 1~Gyr, respectively.   In this model, the sole parameter is $F_{\rm MID}$, the fraction of cluster disrupted after each decade in age, with $F_{\rm MID} = 0.9$ being the standard value.

A variant of this model has been put forward by Elmegreen \& Hunter~(2010) and Kruijssen et al.~(2011a), where clusters are destroyed through strong interactions with the hierarchical interstellar medium, a process which may be mass independent (if the shocks are strong enough, otherwise it will be mass dependent) but should depend strongly on the ambient density of the gas (i.e. environment).

\subsection{M83 as a test case}

In order to differentiate between the two models discussed above, we have carried out a cluster population study in M83, based on the early release science observations with WFC3 on {\it HST}.  The observations, source selection and fitting methods are discussed in detail in Bastian et al.~(2011) and Bastian et al.~(in prep).  The main difference between the dataset used here, and previous studies of cluster disruption, is that an attempt was made in the present case to separate clusters from associations (i.e. only objects that satisfy the criteria that they are older than a current crossing time are considered; Gieles \& Portegies Zwart~2011).  This was done through visual inspection of all detected candidates (carried out in the usual automated way) and only resolved, centrally concentrated and symmetric sources were considered.  See Silva-Villa \& Larsen~(2010; 2011) for further discussions on the effects of visual inspection.  Additionally, we have not included the inner $450$~pc of the galaxy, due to the significantly lower detection limit in this region, high and patchy extinction, and uncertain star-formation history (SFH).  Outside this region, the SFH of M83 appears to have been largely constant over the past 100~Myr (Silva-Villa \& Larsen 2011).  Finally, we limit our analysis to clusters with masses in excess of $5\times10^3$\msun, in order to limit stochastic sampling effects (e.g., Fouesneau \& Lan{\c c}on~2010; Silva-Villa \& Larsen~2011).

\subsection{Environmental dependence}

The HST WFC3 observations cover a wide range of galacticentric radii, extending from the nucleus to $\sim6$~kpc.  Hence, a wide range of environments are sampled by the data.  We split our sample into an inner and outer region, and search for differences in the observed properties.  Due to the lower surface density of GMCs and weaker tidal field in the outer regions of the galaxy, the MDD scenario predicts that clusters should live longer in the outer regions, hence that the average age should be older for increasing galactocentric radii.  The MID scenario, on the other hand, predicts that cluster lifetimes are independent of location, resulting in a 'universal age/mass distribution', hence the inner/outer fields should show the same distribution.

\begin{figure}
\center
\includegraphics[width=12cm,angle=0,clip=true]{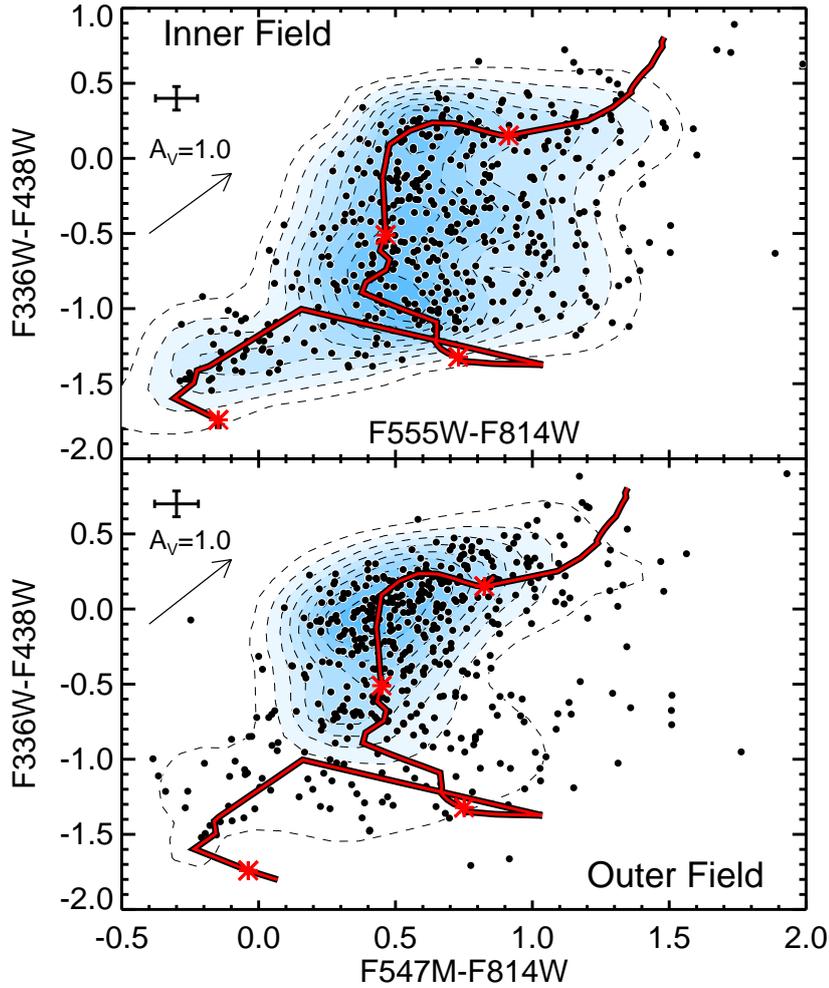} 
\caption{\label{fig:colour_contours} (from Bastian et al.~2011) Colour-colour diagrams for the inner (upper panel) and outer (lower panel) fields.  Each cluster is shown as a filled circle, while the contours denote the number density of points at that colour.  The (red) solid lines are the SSP model tracks used in Adamo et al.~(2010).  For reference the models are marked with asterisks at 1, 10, 100 \& 1000~Myr, from lower-left to upper-right.  Note the differences in the distributions, with a significantly higher fraction of clusters in the outer field having colours consistent with being older.   The average error in colour for our cluster sample is shown in the upper left of each panel, and the extinction vector is also shown.
}
\end{figure}

We use two methods to distinguish between the theories.  The first is (largely) independent of adopting specific simple stellar population (SSP) models.  In Fig.~\ref{fig:colour_contours} we show colour-colour diagrams of clusters in both of the fields.  Filled circles represent individual clusters while the contours show the number density of clusters at that position.  The red lines are SSP model tracks (see Adamo et al.~2010 and Zackrisson et al.~2011), with the astericks labelling ages of 1, 10, 100 \& 1000~Myr, from lower-left to upper-right.  It is clear that the colour distribution of sources is different between the two fields, with the peak of the distribution happening at redder $F336W - F438W$ (essentially $U - B$) colours in the outer field.  The redder $U-B$ colour (and similar $V-I$ colour) suggests an older average age of clusters in the outer field.  Already here, it is clear that the distributions are different, inconsistent with the MID prediction.  We note that extinction cannot explain the observed differences as 1) a lower average extinction in the outer field is expected and 2) the difference between the fields is mostly in the $U-B$ colour, whereas the $V-I$ colour distribution is largely the same.

As a second test between MID and MDD, we derived the age and mass of all the clusters in our sample by comparing the observed colours/magnitudes to SSP models (see Bastian et al.~2011 for details on the method and models used).  We then look at the mass distribution of clusters in three separate age bins, normalised to the linear age range covered.  This is the same method used by Chandar et al.~(2010b) to study cluster disruption in the inner field of M83.  If cluster disruption (or cluster mass loss) was negligible, then the three bins would be expected to lie (nearly) on top of each other.  If, on the other hand, cluster disruption was independent of mass and that 90\% of clusters were disrupted (or each cluster lost 90\% of its mass) then the mass function at each age should be a factor of 10 below the previous age.

\begin{figure}
\center
\includegraphics[width=12cm,angle=0,clip=true]{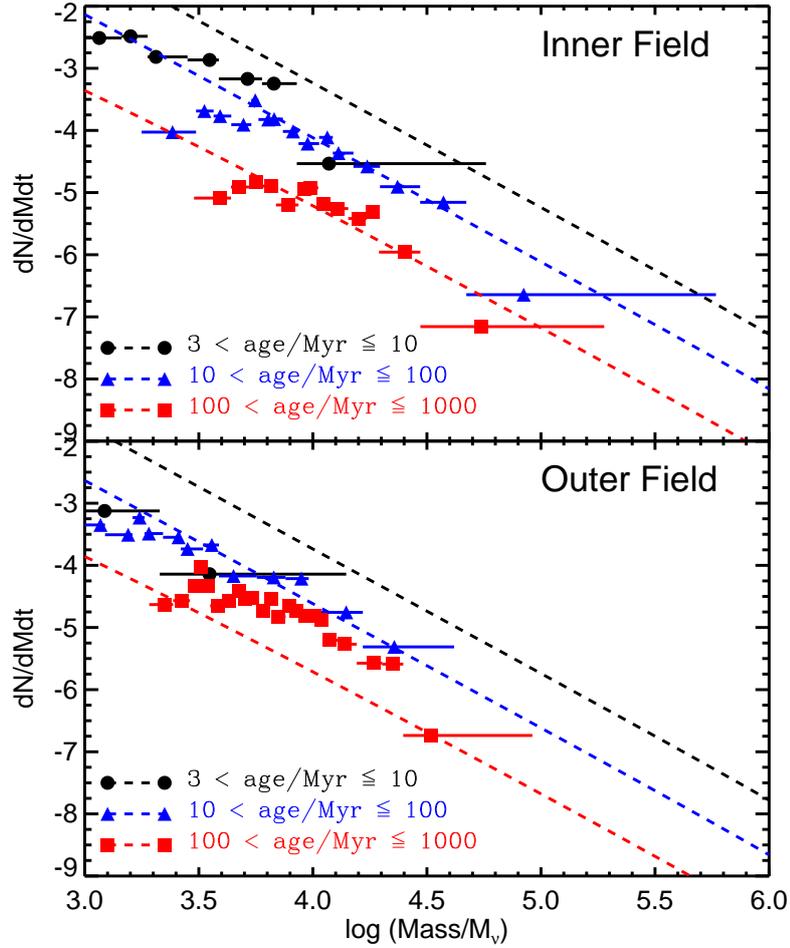} 
\caption{\label{fig:dndmdt} (from Bastian et al.~2011) The mass distributions for three age bins (normalised to the linear age range considered) for the inner (top panel) and outer (bottom panel) fields.  The points represent the median mass in each mass bin, while the horizontal lines show the mass range covered by the bin.  The dashed lines show the predicted ``universal" cluster distributions of the mass independent disruption scenario ($\fmid=0.9$) normalised to fit the 10-100~Myr distributions.  We have used bins with equal numbers of clusters (10 per bin).  The distributions in the upper and lower panels are clearly different, suggesting that disruption has been much less pronounced in the outer field than in the inner field.
}
\end{figure}

In Fig.~\ref{fig:dndmdt} we show the observed distributions. The first age bin (3-10~Myr) has the caveat that some clusters may have been missed due to the elimination of associations from the sample (discussed in detail in Bastian et al.~in prep).  The upper panel (ignoring the first age bin) agrees well with the results of Chandar et al.~(2010b), who used the same data, but had different selection criteria\footnote{This implies that after the first $\sim10$~Myr, clusters are fairly well defined objects and that different selection algorithms will generally do a good job in finding them.  Before this age, the samples will be heavily influenced by the selection criteria used.}.  In the inner field, the 10-100~Myr age bin is significantly higher than the 100-1000~Myr bin (roughly by a factor of 10), indicating a high level of disruption.  In the outer field, however, the mass distributions (again focussing only on the last two age bins) lie nearly on top of each other, suggesting that little or no disruption has been acting on the population.

{\it The observed differences in the age/mass distributions between the inner and outer field indicate that disruption has been much more significant in the inner regions of M83 than in the outer regions.  This is inconsistent with the MID scenario (where a universal distribution is expected, independent of environment) and in good agreement with predictions of the MDD paradigm.}  It is also qualitatively consistent with the environmentally dependent MID scenario of Elmegreen \& Hunter~(2010) and Kruijssen et al.~(2011a).

\subsection{Mass dependence}

The role of mass in the cluster disruption process is more difficult to discern than environmental influences.  The reason is that the form of the cluster mass function plays a significant role.  For example, Chandar et al.~(2010) have shown that if the cluster mass function is a pure power-law with an index of -2, then the MDD scenario is not able to reproduce the observed distribution of clusters in the inner field of M83.  The reason is simple, in the MDD scenario, high mass clusters, with masses above e.g., $10^6$\msun, are not expected to lose much mass.  This means that the predicted models for mass distributions (such as that shown in Fig.~\ref{fig:dndmdt}) should all converge at high masses.  However, if the mass function has a truncation at high masses, then the models no longer converge at the high mass end.

Evidence for a truncation at the high mass end has been seen in the luminosity function of clusters (Gieles et al. 2006), the relation between the magnitude brightest cluster in a population and the host galaxies star-formation rate (Bastian~2008), in the relation between the age and brightness of the brightest (and fifth brightest) clusters in a population (Larsen~2009), as well as through direct fitting of the mass function clusters within a galaxy (Larsen~2009; Bastian et al.~in prep).

Hence, in order to see if there is a dependence on mass in the disruption process, and distinguish between the MID and MDD scenarios, we must fit on the full age/mass distributions for three parameters.  The parameters are: $t_{\rm 4}$ - the disruption timescale of a $10^4$~\msun\ cluster in the MDD scenario; $F_{\rm MID}$ - the fraction of cluster disrupted every decade in age in the MID scenario; and $M_{c}$ - the characteristic mass in the cluster mass function, above which the distribution drops exponentially (we approximate the mass function as a Schechter function).

We adopt a value of $\gamma=0.62$ (in the MDD scenario the disruption timescale is $t_{\rm dis} = t_{\rm 4} (M/10{^4}\msun)^{\gamma}$), which was found empirically by Boutloukos and Lamers (2003) and Lamers et al. (2005) and agrees with N-body simulations
for clusters with an initial density distribution of a King profile of W$_0$=5 (Baumgardt and Makino, 2003;  Gieles et al. 2004).  Models with W$_0$=7 have $\gamma=0.70$ (Lamers et al 2010).

\begin{figure}
\center
\includegraphics[width=7.3cm,angle=0,clip=true]{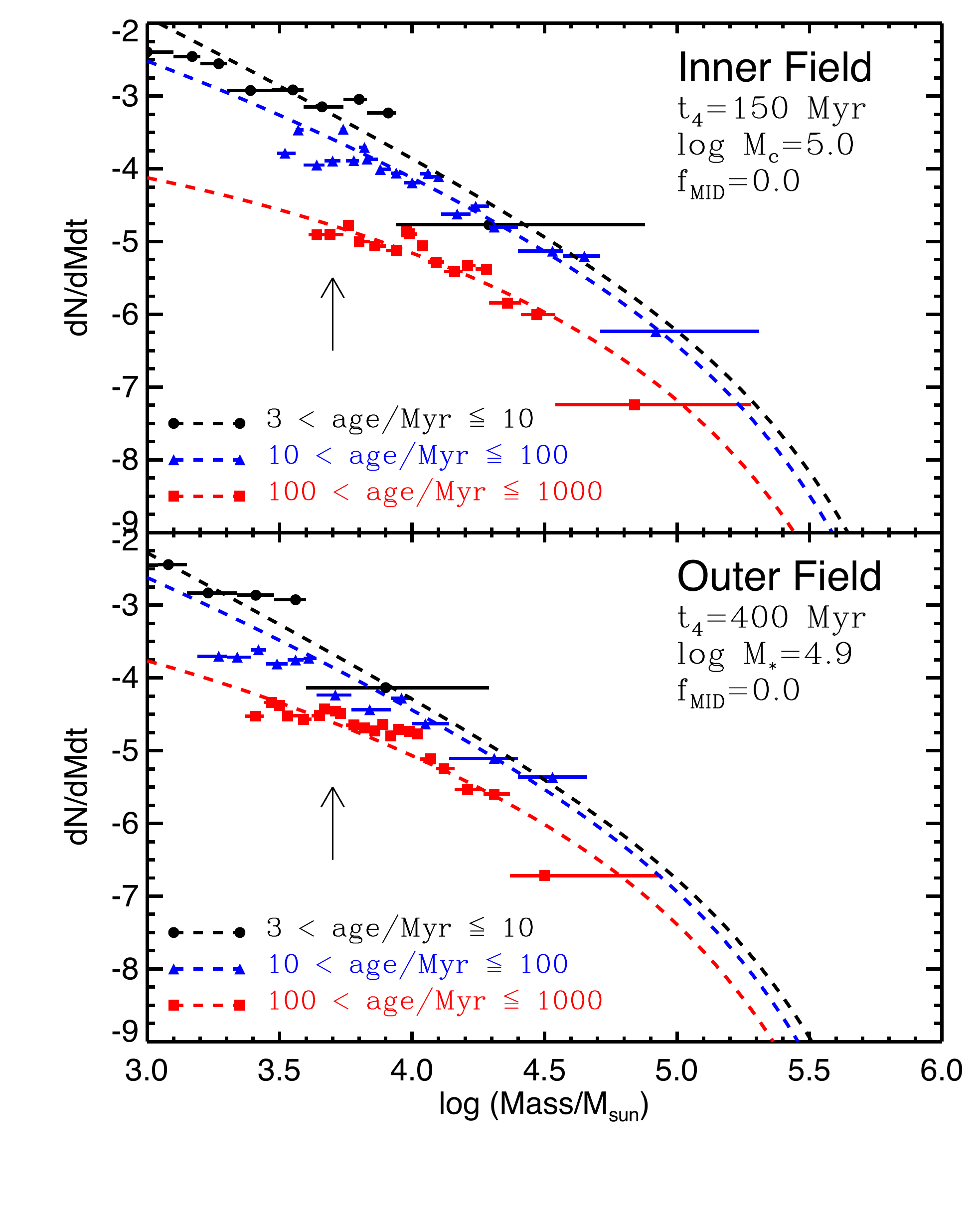} 
\includegraphics[width=7.3cm,angle=0,clip=true]{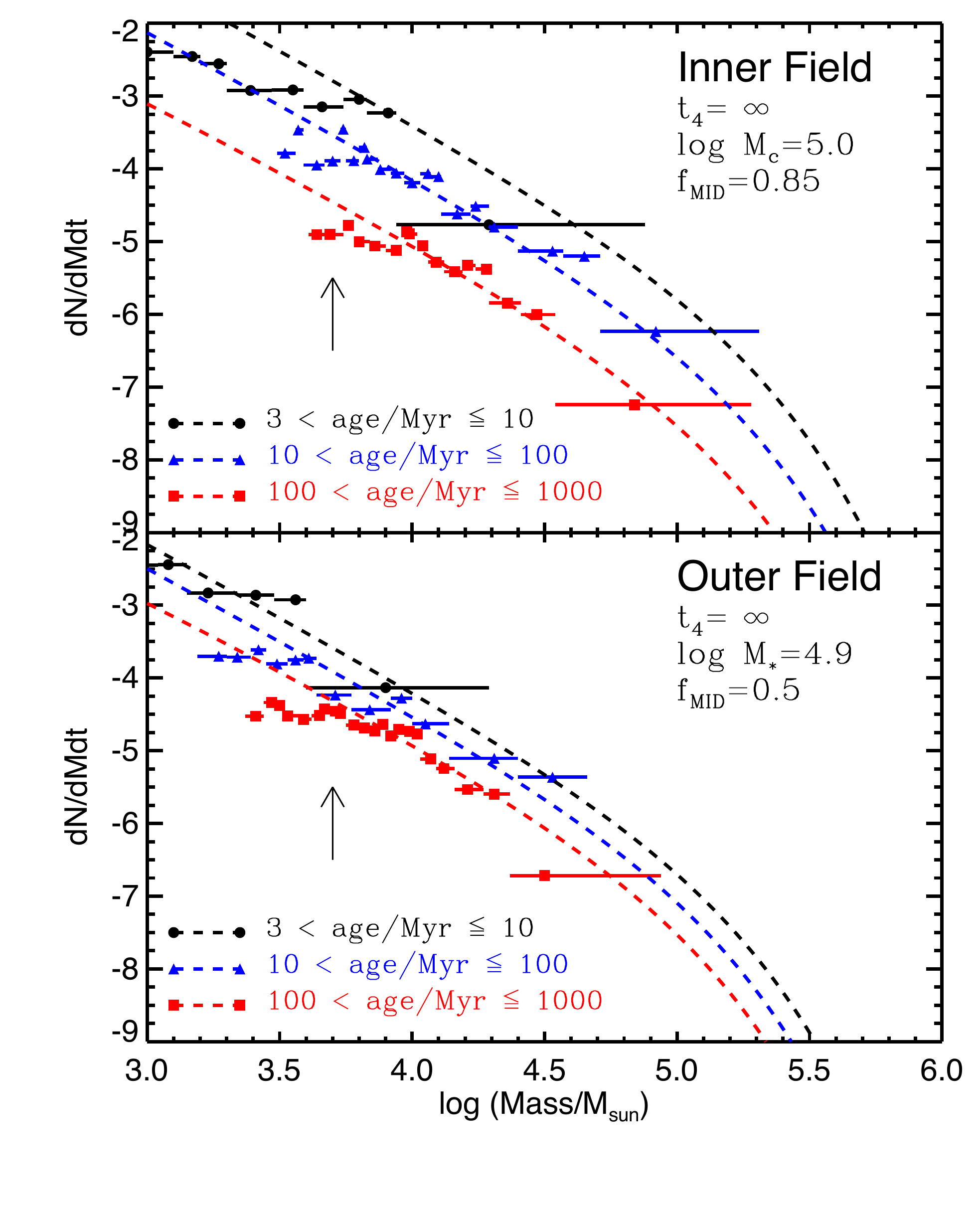} 

\caption{\label{fig:dndmdt_mod} (from Bastian et al.~in prep) The same as Fig.~\ref{fig:dndmdt} except now the best fitting pure MDD (left panels) and MID (right panels) models are shown and a Schechter mass function is used.  Note that both models can reproduce the data.  However, in the MID case, different values of $F_{\rm MID}$ are needed, 0.85 in the inner field and 0.5 in the outer field.  The arrows denote the lower mass limit adopted.}
\end{figure}

As a first step we apply the pure MID and MDD models to the same M83 data shown in Fig.~\ref{fig:dndmdt} and the best fitting models are shown in Fig.~\ref{fig:dndmdt_mod}.  The left panels show the results for the pure MDD case (i.e. $F_{\rm MID} = 0$) and the right panels show the best fit for the MID case (i.e. $t_{\rm 4} = \infty$).  We see that both the MID and MDD models can reproduce the observed distributions.  However, in the MID case, different values of $F_{\rm MID}$ are needed for the inner and outer field, again suggesting non-universal age/mass distributions.

\begin{figure}
\center
\includegraphics[width=15.3cm,angle=0,clip=true]{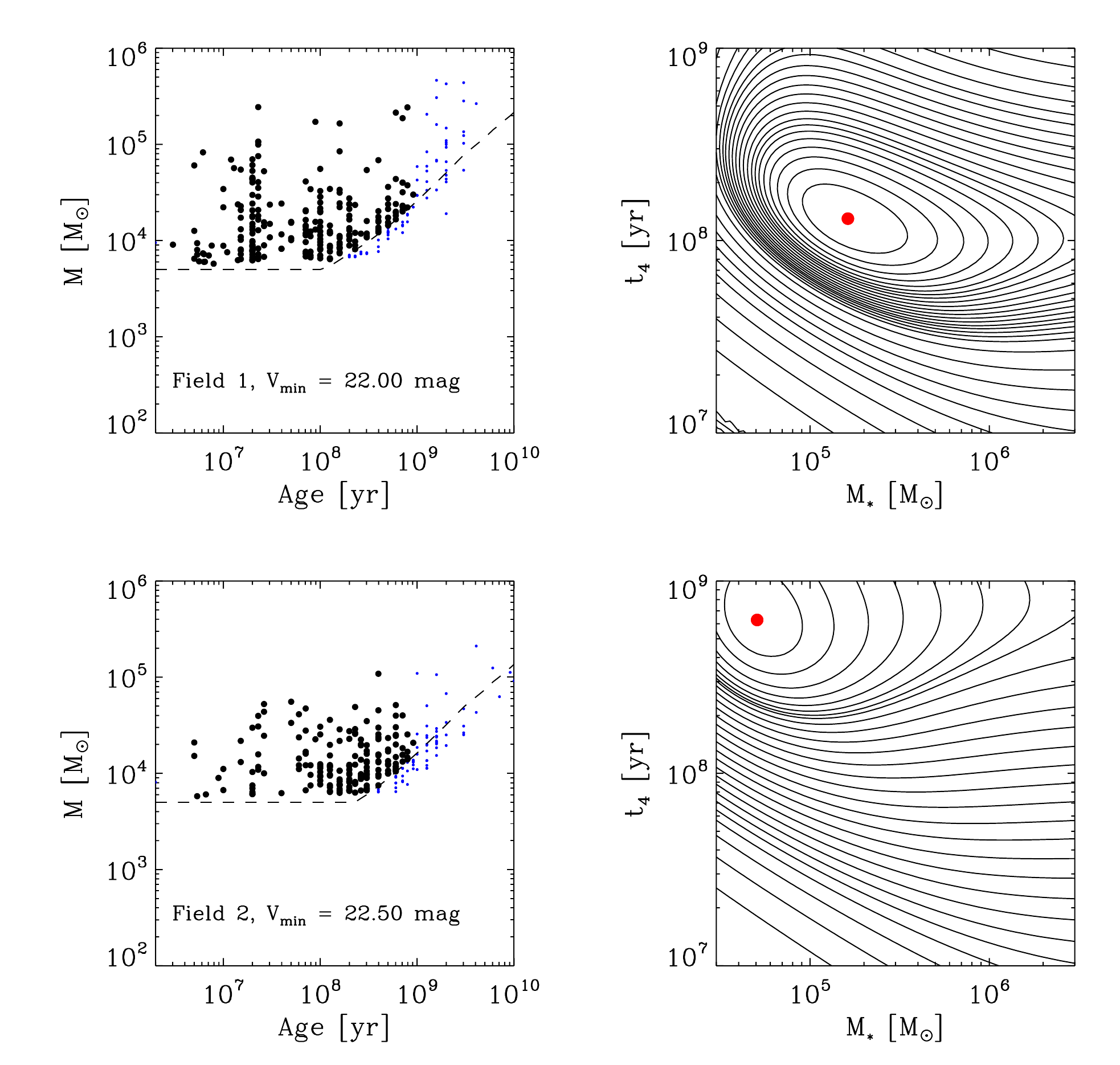} 
\caption{\label{fig:maxlike} (from Bastian et al.~2011b) {\bf Left panels:} The age/mass diagrams for clusters in the inner (top) and outer (bottom) fields.  Solid circles denote clusters that were used in the population fitting.  The (blue) points represent clusters not used.  The dashed line represents the surveys completeness limits, mass limited at young ages ($5\times10^3$\msun) and luminosity limited at older ages (the adopted limits are shown in the panels).  {\bf Right panels:}  The results of a maximum likelihood fitting of each population, assuming only mass dependent disruption and a Schechter mass function.  The solid (red) dotes shows the best fitting model.}
\end{figure}

Finally, in order to avoid any artefacts of the binning from affecting the results, we have also carried out a maximum likelihood fitting analysis of the cluster population.  Since the MID model advocated by Whitmore et al.~(2007) and Fall et al.~(2009) (that $F_{\rm MID}$ is universal) does not fit the data, we have limited the fit to only the MDD case.  The results are shown in Fig.~\ref{fig:maxlike}.  The full details of the fitting will be presented in Bastian et al.~(in prep), where MID will also be included as a free parameter.

The results indicate that cluster disruption has been much more rapid in the inner field ($t_{\rm 4} \sim 130$~Myr) than the outer field ($t_{\rm 4} \sim 600$~Myr).  The difference between the two fields is generally consistent with the predictions of the MDD theory (Lamers et al.~2010).  Additionally, we see that both fields require a truncation in the cluster mass function, with $M_{\rm c} \sim 10^5$\msun, being slightly larger in the inner field than the outer field.  This is in good agreement with the estimated value of $M_{\rm c}$, using independent methods, for spiral galaxies (Gieles et al.~2006; Larsen~2009) and for the LMC (Maschberger \& Kroupa~2009).

\section{Conclusions}

We have reviewed the standard paradigm of cluster formation and early dissolution.  The role of gas expulsion in the disruption process is strongly dependent on the initial conditions of the system.  Sub-virial and hierarchical stellar groups can evolve in such a way that gas expulsion has little or not effect on their evolution.  Additionally, recent studies of the spatial distribution of YSOs in the solar neighbourhood suggest that only a small fraction of stars form in groups that would be associated with ``clusters", with the majority formed in low density groups/associations that will disperse into the background without the need of gas expulsion.  Studies of the dynamics of resolved massive clusters in the Galaxy and LMC have found low velocity dispersions, suggesting that previous studies of unresolved young clusters (that found high velocity dispersions) were affected by high-mass binaries.  These resolved massive clusters appear to be in a stable (or even sub-virial) configuration, even at very young ages, meaning that gas expulsion has had little or no effect on their evolution.  {\it Hence, it appears that the effect of gas expulsion on young clusters and the role of infant mortality in cluster populations has been overestimated.}

After a cluster passes from an embedded to an exposed phase, a variety of processes will act upon the cluster which may lead to its disruption.  We have reviewed the two main empirical disruption laws of clusters, {\it mass dependent disruption} (MDD) and {\it mass independent disruption} (MID).  In order to test the two models, we have used new {\it HST} observations of the cluster population in M83.  The properties of the cluster population clearly change as a function of galactocentric distance, with clusters in the outer portion of the galaxy having higher mean ages.  This is in agreement with the MDD scenario, which predicts that cluster disruption should depend on the ambient environment.  While the role of mass in the cluster disruption process is more difficult to assess, due to the added parameter of the shape of the cluster mass function in the fitting analysis, we find evidence that higher mass clusters live longer.  However, both the MDD and MID models can fit the data if all three main parameters ($F_{\rm MID}$, $t_{\rm 4}$, and $M_{\rm c}$) are allowed to vary.  Finally, the cluster population of M83 requires a truncation in the mass function at $\sim10^5$\msun, which is similar to that found for other galaxies, using independent techniques.

%
%
\small  
%
\section*{Acknowledgments}   
%
We thank the organisers for a very lively and stimulating workshop.  In additional, I am indebted to my collaborators, both for their insights and discussions as well as their help on the projects presented here.  Namely, I would like to thank: Angela Adamo, Mark Gieles, Eli Bressert, Rob Gutermuth, Henny Lamers, Esteban Silva-Villa, Soeren Larsen, Simon Goodwin, Diederik Kruijssen, Linda Smith, Iraklis Konstantopoulos, and Erik Zackrisson.
%

%


\end{document}